\documentclass[conference]{IEEEtran}
\IEEEoverridecommandlockouts
\usepackage{cite}
\usepackage{balance}
\usepackage{amsmath,amssymb,amsfonts}
\usepackage{algorithmic}
\usepackage{graphicx}
\usepackage{textcomp}
\usepackage{xcolor}
\usepackage[T1]{fontenc}
\usepackage[utf8]{inputenc}
\usepackage{listings}
\usepackage{tikz}
\usepackage{bbm}
\usetikzlibrary {fadings}

\usepackage{color}

\definecolor{keywordcolor}{rgb}{0.7, 0.1, 0.1}   
\definecolor{tacticcolor}{rgb}{0.0, 0.1, 0.6}    
\definecolor{commentcolor}{rgb}{0.4, 0.4, 0.4}   
\definecolor{symbolcolor}{rgb}{0.0, 0.1, 0.6}    
\definecolor{sortcolor}{rgb}{0.1, 0.5, 0.1}      
\definecolor{attributecolor}{rgb}{0.7, 0.1, 0.1} 

\newtheorem{definition}{Definition}

\newcommand{\todo}[1]{}
\newcommand{\spacetune}[1]{#1}

\usepackage{graphicx}
\def\BibTeX{{\rm B\kern-.05em{\sc i\kern-.025em b}\kern-.08em
    T\kern-.1667em\lower.7ex\hbox{E}\kern-.125emX}}
\begin{document}

\title{Structuring Definitions in Mathematical Libraries\\
{}
}

\author{\IEEEauthorblockN{Alena Gusakov}
\IEEEauthorblockA{\textit{Cheriton School of Computer Science} \\
\textit{University of Waterloo}\\
Waterloo, Canada\\
\texttt{agusakov@uwaterloo.ca}}
\and
\IEEEauthorblockN{Peter Nelson}
\IEEEauthorblockA{\textit{Dept of Combinatorics and Optimization} \\
\textit{University of Waterloo}\\
Waterloo, Canada\\
\texttt{apnelson@uwaterloo.ca}}
\and
\IEEEauthorblockN{Stephen M. Watt}
\IEEEauthorblockA{\textit{Cheriton School of Computer Science} \\
\textit{University of Waterloo}\\
Waterloo, Canada\\
\texttt{smwatt@uwaterloo.ca}}
}

\maketitle

\begin{abstract}
Codifying mathematical theories in a proof assistant or computer algebra system is a challenging task, of which the most difficult part is, counterintuitively, structuring definitions. This results in a steep learning curve for new users and slow progress in formalizing even undergraduate level mathematics. There are many considerations one has to make, such as level of generality, readability, and ease of use in the type system, and there are typically multiple equivalent or related definitions from which to choose. Often, a definition that is ultimately selected for formalization is settled on after a lengthy trial and error process. This process involves testing potential definitions for usability by formalizing standard theorems about them, and weeding out the definitions that are unwieldy. 

Inclusion of a formal definition in a centralized community-run mathematical library is typically an indication that the definition is ``good.'' For this reason, in this survey, we make some observations about what makes a definition ``good,'' and examine several case studies of the refining process for definitions that have ultimately been added to the Lean Theorem Prover community-run mathematical library, \texttt{mathlib}. We observe that some of the difficulties are shared with the design of libraries for computer algebra systems, and give examples of related issues in that context.
\end{abstract}


\section{Introduction}
Type theoretic proof assistants for mathematics formalization can have a steep learning curve for new users. 
A similar situation exists for computer algebra systems with elaborate type hierarchies.  
A definition in a proof assistant is an interface for the properties and data of the mathematical object written in a specific formal logic system. For this reason, developing definitions in a formal language calls for a robust knowledge of how the underlying logic system will govern the interactions between different definitions, notation, automated or implicit inference, and so on. 

It can take a great deal of trial and error to settle on an acceptable definition or hierarchy of definitions. This is often done via a cyclic process of writing a definition, developing a theory around it as a proof of concept, and going back to adjust or rewrite the definition as appropriate. In some cases one may be able to construct a hierarchy of definitions in which properties can be inherited, but in other cases the hierarchy is flat and one must define various canonical maps between definitions. In some cases one might find the need to define a single object in several equivalent ways.

In both the proof assistant and computer algebra settings, the choice of definitions has far-reaching consequences on library structure, including understandibility, elegance of expression, composibility of components, automation, ease of specialization and generalization, and efficiency of code compilation and execution.

It is difficult for the authors to create specific guidelines for how to structure definitions or hierarchies, as one typically develops an intuition for how a type system behaves via direct experience with it, familiarity with its rules, and exposure to examples. This paper attempts to provide some examples and analyses.


\section{Dependent Type Theory in Lean}

There are many different grammars that can be used to expressed mathematical ideas: dependent type theory, set theory, univalent type theory, and so on. Each logic system has pros and cons when it comes to formalization, and this is a source of much debate within and between different proof assistant communities. One feature of dependent type theory is that it requires a massive amount of rigor. This often bogs users down with tedious details, but this tedium also prevents users from unintentionally writing or proving nonsensical statements --- for instance, proof assistants that use set theory often require an additional layer of rigor 
(\textit{e.g.} Mizar requires an additional layer of type theory in order to avoid this \cite{Bauer}).
For this reason, in this paper we focus on the Lean Theorem Prover, a proof assistant that uses dependent type theory.

In dependent type theory, every object is a term or type, and every term has a unique type. The strictness of this typing system, in particular the uniqueness of a type for a given object, means that we have to be careful when expressing and proving mathematical equality. There are two ``main'' types of equality that we are concerned with in Lean: definitional and propositional.

When we say that two things are definitionally equal, we mean that the type system can recognize that they are the same via some automatic reductions. By contrast, when we say two things are propositionally equal, what we mean is that the user has to supply a proof.

Propositional equalities introduce overhead, so users who develop formal definitions in a proof assistant strive to minimize or eliminate propositional equality in favor of definitional equality, especially for ``trivial" or simple expressions. Most of the design decisions we discuss in this paper are made with this intention.

Dependent types, subtypes and type coercions are a source of type theoretic difficulties. Dependent types interfere with type equality, e.g. if we have some indexing function \lstinline{A : ℕ → Type*}, then \lstinline{A (1 + 0)} will not be recognized automatically as the same as \lstinline{A (0 + 1)} \footnote{even if \lstinline{0 + 1} is defined as we typically understand it}. We can prove propositional equality between the two, but since the types are not definitionally equal we must manually coerce between them.

Because of uniqueness of types, type coercion is a necessary feature for being able to express any sort of hierarchy between mathematical objects in a type theory system. However, while Lean provides an automatable framework for dealing with type coercions, it can get quite difficult as the layers of coercions add up. One has to carefully consider how to use types, subtypes, and coercions within definitions and expressions, as well as keep track of automatic  coercions.

\section{Developing Definitions}

There are examples of mathematical theories that lend themselves nicely to formalization, such as undergraduate-level algebra. Mathematics in Lean \cite{avigad2025mathematics} provides an extensive explanation of the algebra hierarchy in \texttt{Mathlib}, so we will not retread those steps. Instead, in this section we examine some examples of difficulties with formalizing other mathematical theories. 

When formalizing definitions, it is important to find a balance between code readability, correctness, and ease of use via minimizing type theoretic annoyances (subtypes, coercion, dependent types, propositional equality). Since there are many ways of expressing the same definition, this becomes a source of much discussion. Often the type theoretic features are unavoidable when we express certain mathematical notions, and users collaboratively invent various workarounds\footnote{e.g. in the Zulip chat server}. In addition, one has to ``stress test'' a chosen definition by formalizing theorems about it, as often the type theoretic difficulties only become apparent as a user attempts to prove statements.

We will explore a few cases of this process. Some examples come from existing published work, so we attempt to provide a high-level overview of the papers' main points. By contrast, the last two subsections on matroid and graph theory formalization contain unpublished work, so we provide an in-depth exploration of our observations and workarounds. 

\subsection{Continuous Functional Calculi}

The formalization efforts in \cite{Dedecker2025} focus on defining continuous functional calculi in a way that allows straightforward use of commonly used tools in the theory as well as nice generalizations. In this section, we examine a subset of these efforts while omitting mathematical details to focus on interactions at the type theoretic level. 

The authors' first, and mathematically correct, attempt at formalizing the definition is as follows:

\begin{lstlisting}
/-- The ⋆-isomorphism between the continuous functions on the spectrum of `a` and the (unital) C⋆-subalgebra generated by `a`. -/
def continuousFunctionalCalculus 
{A : Type*} [CStarAlgebra A]
(a : A) [IsStarNormal a] :
  C(spectrum ℂ a, ℂ) ≃⋆ₐ[ℂ] StarAlgebra.elemental ℂ a
  -- implementation details omitted
\end{lstlisting}
Without delving into the mathematical details, we can break down the following source of dependent type problems: \lstinline{C(spectrum ℂ a, ℂ)} is a type of functions \lstinline{↑(spectrum ℂ a) → ℂ} with a bundled condition of continuity. This means that, if we have some function \lstinline{f : ℂ → ℂ} that we know is continuous on the spectrum of \lstinline{a}, we cannot plug it into the isomorphism directly because the type of domain \lstinline{ℂ} does not match with expected domain type \lstinline{↑(spectrum ℂ a)}. Instead, we must plug in its restriction to \lstinline{↑(spectrum ℂ a)} as well as the proof of continuity. This significantly hampers readability, and because \lstinline{↑(spectrum ℂ a)} is a type that depends on \lstinline{a}, we will encounter additional friction with type equalities\footnote{e.g. in the case of \lstinline{b : A} with propositional equality \lstinline{a = b}, the \lstinline{IsStarNormal} instance on \lstinline{a} would prevent us from directly replacing \lstinline{a} with \lstinline{b}.}.



The authors create an interface to avoid this problem using a "junk value" approach\footnote{This is design a pattern that appears throughout \texttt{Mathlib}, as in the case of the multiplicative inverse of \lstinline{0 : ℝ} defined as \lstinline{(0 : ℝ)⁻¹ := 0}, which appears in other notable proof assistant libraries.}. Here they relax the domain and codomain of the definition to their parent type \lstinline{A}, and accept parameter \lstinline{(f : ℂ → ℂ)} instead of a function with domain type \lstinline{↑(spectrum ℂ a)}. The implementation then uses an \lstinline{if then else} function that plugs in the arguments to \lstinline{continuousFunctionalCalculus} if the restriction is continuous on the spectrum of $a$ (and if \lstinline{IsStarNormal a}), and outputs \lstinline{0} otherwise:

\begin{lstlisting}
def cfc (f : ℂ → ℂ) (a : A) : A :=
  if h : IsStarNormal a 
    ∧ Continuous ((spectrum ℂ a).restrict f)
  then
    letI := h.1 -- h.1 gives [IsStarNormal a]
    continuousFunctionalCalculus a 
        ⟨(spectrum ℂ a).restrict f, h.2⟩
  else 0
\end{lstlisting}
This loses some mathematical content, such as the fact that the map is an isomorphism. However, these facts can be derived as separate lemmas since the definition of \lstinline{cfc} relies on \lstinline{continuousFunctionalCalculus}, and this is outweighed by the ease of use\footnote{For an example of this nice property, consider how one would express the statement \lstinline{cfc (g ∘ f) a = cfc g (cfc f a)} using the original definition of \lstinline{continuousFunctionalCalculus}.}.

This iteration does not generalize easily to domains besides \lstinline{ℂ}, as there are domains where a \lstinline{⋆}-isomorphism with the same properties exist with different codomains. In order to avoid creating a new definition for every possible domain, the authors convert the definition into a typeclass on parameter \lstinline{(R : Type*)} with weakened versions of the required instances. The authors then provide additional conditions that allow us to re-derive the isomorphism as appropriate such as \lstinline{map_id}: 

\begin{lstlisting}
class ContinuousFunctionalCalculus 
(R : Type*) {A : Type*} 
[/-required instances-/] (a : A) where
toStarAlgHom : C(spectrum R a, R) →⋆ₐ[R] A
map_id : 
    toStarAlgHom (.restrict (spectrum R a) (.id R)) = a
-- some fields omitted 
\end{lstlisting}
The authors also relax the \lstinline{CStarAlgebra} instance to allow for the definition to be applied on the type \lstinline{Matrix n n ℝ} of \lstinline{n} by \lstinline{n} square matrices over \lstinline{ℝ}, or type \lstinline{Matrix n n ℂ} of square matrices over \lstinline{ℂ}, and much of the rest of \cite{Dedecker2025} describes these further generalizations.

Now, Lean would not be able to automatically derive this typeclass instance, as any argument \lstinline{(a : A)} would need to be accompanied by proofs that it satisfies the conditions. The big shift in perspective that the property needed to satisfy the definition is the same for every element of the algebra. Thus, we can view \lstinline{ContinuousFunctionalCalculus} as a property of the algebra instead of its elements. This can be expressed using the \lstinline{outParam} as described in \cite{avigad2025mathematics} - this allows us to supply a predicate \lstinline{p} on type \lstinline{A}, with a proof that any element that satisfies the predicate \lstinline{p} satisfies the conditions in the definition. This leads to the following version:

\begin{lstlisting}
class ContinuousFunctionalCalculus 
(R : Type*) {A : Type*} 
(p : outParam (A → Type*))
[/-required instances-/] where
toStarAlgHom {a} (ha : p a) : 
    C(spectrum R a, R) →⋆ₐ[R] A
map_id {a} (ha : p a) : toStarAlgHom ha (.restrict (spectrum R a) (.id R)) = a
-- some fields omitted
predicate_preserving {a} (ha : p a) 
    (f : C(spectrum R a, R)) : 
        p (toStarAlgHom ha f)
\end{lstlisting}
Note that the authors add field \lstinline{predicate_preserving}. Since we have an abstract predicate \lstinline{p}, this field ensures that our choice of \lstinline{p} allows us to compose the definition with itself without having to prove compatibility each time.

We describe one final push to facilitate Lean's automatic typeclass inference. The field \lstinline{toStarAlgHom} with type \lstinline{C(spectrum R a, R) →⋆ₐ[R] A} is a potential source of type equality problems in the case of \lstinline{Matrix n n ℂ}, as it is possible to derive two distinct instances of the type class with \lstinline{toStarAlgHom} fields that are not definitionally equal. Since the definition applies to \lstinline{Matrix n n ℂ}, this is something that must be addressed. The authors make the observation that the \lstinline{⋆}-morphism is unique for domains such as \lstinline{ℝ} or \lstinline{ℂ} due to the Stone-Weierstrauss Theorem, so they are able to abstract function \lstinline{toStarAlgHom} to condition \lstinline{exists_cfc_of_predicate}, which merely states that a \lstinline{⋆}-morphism exists. This takes advantage of the fact that Lean's dependent type theory has proof irrelevance, i.e. all proofs of a \lstinline{Prop} are definitionally equal to each other, allowing this entire typeclass to be turned into a \lstinline{Prop}. 

\begin{lstlisting}
class ContinuousFunctionalCalculus 
(R : Type*) {A : Type*} 
(p : outParam (A → Prop))
[/-required instances-/] : Prop where
-- some fields omitted
exists_cfc_of_predicate : ∀ a, p a → 
-- the desired ⋆-morphism
∃ φ : C(spectrum R a, R) →⋆ₐ[R] A,
    -- restatement of map_id:
    φ ((ContinuousMap.id R).restrict <| spectrum R a) = a 
    ∧ -- other conditions omitted
\end{lstlisting}
Then, the \lstinline{⋆}-morphism can be extracted in a new definition, and \lstinline{cfc} can be refactored for this new definition in order to avoid the sub- and dependent type problems from earlier.

\subsection{Graded Rings}

Graded rings are another context in which dependent types are a source of problems, as documented in \cite{Wieser2022}. Without going too deeply into the definitions, we will view the formalization from the perspective of how it is implemented with respect to types.

A graded algebra over a ring $R$ is a class of additive subgroups $A_0, A_1, A_2, \dots \subseteq R$ where $R$ is the direct sum of the $A_i$'s, every element of $R$ can be expressed uniquely as a sum of one element from each $A_i$, and $A_iA_j \subseteq A_{i + j}$ for all $i, j$. A classic example is $R[X]$, the polynomial ring over $R$. The $A_i$ are the additive subgroups consisting of degree $i$ polynomials. 

With this definition, we can immediately see a source of dependent type theory problems: The subgroups' indexing. We define the indexing function as \lstinline{A : ι → Type*}, where \lstinline{ι} is our indexing type. Then since groups (and subgroups) are a type in \texttt{Mathlib}, that means additive group \lstinline{A i} is a type that depends on the index \lstinline{i}. 

The indexing function plays an important role in the theory of graded rings, as the additive structure of the indexing type of must be reflected in the multiplicative structure of the algebra. If \lstinline{ℕ} is the indexing type, we have associativity of addition. However, even expressing corresponding multiplicative associativity proves difficult: if we have \lstinline{x : A i}, \lstinline{y : A j}, and \lstinline{z : A k}, then \lstinline{(x * y) * z = x * (y * z)} will immediately give us an error because \lstinline{(x * y) * z} has type \lstinline{A ((i + j) + k)} whereas \lstinline{x * (y * z)} has type \lstinline{A (i + (j + k))}, and \lstinline{((i + j) + k) = (i + (j + k))} is propositionally true.

In \cite{Wieser2022}, the authors consider various approaches to this definition. In particular, the authors focus on the behavior of multiplicative associativity, how difficult the approaches were to use, and whether the definitions give us something closer to or further from definitional equality.

The authors narrow the choices down to two, both of which involve Lean's built-in sigma type \lstinline{(Σ i, A i)} of dependent pairs in different ways. These definitions both make multiplicative associativity very easy to prove.

\begin{lstlisting}
class g_semigroup [add_semigroup ι] extends semigroup (Σ i, A i) :=
(fst_mul {i j : ι} (x : A i) (y : A j) :
(⟨_, x⟩ * ⟨_, y⟩ : Σ i, A i).fst = i + j)

class g_semigroup [add_semigroup ι] :=
(mul {i j : ι} : A i → A j → A (i + j))
(mul_assoc {i j k : ι} (x : A i) (y : A j) (z : A k) :
(⟨_, mul (mul x y) z⟩ : Σ i, A i) = ⟨_, mul x (mul y z)⟩)
\end{lstlisting}
The first definition hooks into existing theory about multiplication from \texttt{Mathlib} by casting the elements of distinct types \lstinline{A i} and \lstinline{A j} into the sigma type. However, this has the drawback of the index being equal to \lstinline{i + j} propositionally. The second definition uses a definition of multiplication between distinct types, but does give us definitional equality, so this is the definition the authors ultimately select. 

The authors then redefine \lstinline{mul} as a ring homomorphism 

\subsection{Matroids}




Matroids have many cryptomorphic definitions, and the question of which one to use as the main definition is something that requires some forethought. In order to allow for infinite matroids (thus allowing the proper level of generality), we will consider the following two definitions from \cite{Bruhn2013}.

\begin{definition}
    A \textit{matroid} $M = (E, \mathcal{I})$ is a finite set $E$ with some notion of independence $\mathcal{I}$ defined on subsets of $E$ such that
    \begin{enumerate}
        \item[(I0)] The empty set $\varnothing$ is in $\mathcal{I}$.
        \item[(I1)] \textbf{Monotonicity:} If $X, Y \subseteq E$ such that $X \subseteq Y$ and $Y \in \mathcal{I}$, then $X \in \mathcal{I}$.
        \item[(I2)] \textbf{Augmentation:} If $X, Y \subseteq E$ such that $|X| < |Y|$ and $X, Y \in \mathcal{I}$, then there exists some $y \in Y \backslash X$ such that $X \cup \{y\} \in \mathcal{I}$.
        \item[(I2)] \textbf{Maximality:} For all $X \subseteq E$ and $Z \subseteq X$, such that there exists $I_1 \in \mathcal{I}$ with $Z \subseteq I_1$, there exists a subset $Y \subseteq X$ which is maximal with respect to the property ``there exists $I_2 \in \mathcal{I}$ such that $Y \subseteq I_2$".
    \end{enumerate}
\end{definition}

\begin{definition}
    A matroid $M = (E, \mathcal{B})$ is a set $E$ and a nonempty collection $\mathcal{B}$ of subsets of $E$ such that
    \begin{enumerate}
        \item[(B0)] $\mathcal{B}$ is non-empty
        \item[(B1)] \textbf{Exchange Property:} For all $B_1, B_2 \in \mathcal{B}$, and for all $a \in B_1 \backslash B_2$, there exists $b \in B_2 \backslash B_1$ such that $(B_1 \backslash \{a\}) \cup \{b\} \in \mathcal{B}$ 
        \item[(B2)] \textbf{Maximality:} For all $X \subseteq E(M)$ and $Z \subseteq X$, such that there exists $B_1 \in B$ with $Z \subseteq B_1$, there exists a subset $Y \subseteq X$ which is maximal with respect to the property ``there exists $B_2 \in \mathcal{B}$ such that $Y \subseteq B_2$".
    \end{enumerate}
\end{definition}
We refer to $I \in \mathcal{I}$ as an independent set, or $\mathcal{I}$ as the independence relation on subsets of $E$. Similarly, we refer to $B \in \mathcal{B}$ as a base\footnote{Note that in the matroid theory literature, the term ``basis" is overloaded, as there is also a notion for subsets of the matroid's ground set. In order to make this distinction, we use ``base" for $B \in \mathcal{B}$.}, or $\mathcal{B}$ as the base relation on subsets of $E$.

The intuitive way to think about matroids is as a generalization of vector spaces - linear independence in a module corresponds to independence in a matroid, and a basis for a vector space corresponds to a base of a matroid. For this reason, the closest related definition in \texttt{Mathlib} is \lstinline{Module}. 

However, we cannot look to these definitions for inspiration, as both independent or basis vector sets are derived from properties of a module, and not the other way around as is the case for matroids. What we can do, however, is ensure that our matroid definition works well with these notions.

When defining a matroid, the natural approach would be to define a matroid as having ground type \lstinline{E} with either an independent relation \lstinline{I : Set E → Prop} or a base relation \lstinline{B : Set E → Prop}, with the appropriate additional axioms. This imitates the algebraic hierarchy, and allows us to take advantage of existing lemmas and theorems for \lstinline{ClosureOperator}.

Now, if we consider how the algebra hierarchy works with substructures, we can see that an algebraic object, e.g. \lstinline{Group}, is defined on a type, whereas a substructure \lstinline{Subgroup} is defined on a carrier set of the original type. 

However, this pattern of having the parent object defined on a type and the subobject defined on a set clashes with the fact that we typically perform set operations on the ground set in matroid theory. Furthermore, if we represent the ground set of a matroid as a type, then every time we delete or contract an element, we create a new matroid with a different ground type, which will introduce the need for type coercions every time we go between matroids. 

This leads us to the idea that, instead of having \lstinline{Matroid} defined on a ground type \lstinline{E} with \lstinline{Minor} having a carrier set in the ground type, we collapse the hierarchy so that \lstinline{Matroid} always has a carrier set that belongs to an ambient type. Then if we delete or contract elements from a matroid, we will still obtain a matroid belonging to the same ambient ground type, and do not have to perform type coercions.

This has the drawback of requiring us to perform bookkeeping on set membership, however. If we have independence relation \lstinline{I : Set E → Prop}, we expect that any set that is independent will be a subset of the carrier set, and similar goes for the base relation. One workaround for keeping track of these proofs is by automating as much of them as possible.

While each of the possible axioms for a matroid can be derived from other axioms, the independence and base axioms are particularly closely related. Much like in linear algebra, a base is a maximally independent set, and a set is independent if it is contained in a base. We can use this close relationship to our advantage and introduce both sets of axioms into our main definition of a matroid.

\begin{lstlisting}
structure Matroid (α : Type*) where
  /-- `M` has a ground set `E`. -/
  (E : Set α)
  /-- `M` has a predicate `Base` defining its bases. -/
  (IsBase : Set α → Prop)
  /-- `M` has a predicate `Indep` defining its independent sets. -/
  (Indep : Set α → Prop)
  /-- The `Indep`endent sets are those contained in `Base`s. -/
  (indep_iff' : ∀ ⦃I⦄, Indep I ↔ ∃ B, IsBase B ∧ I ⊆ B)
  /-- There is at least one `Base`. -/
  (exists_isBase : ∃ B, IsBase B)
  /-- For any bases `B`, `B'` and `e ∈ B \ B'`, there is some `f ∈ B' \ B` for which `B-e+f`
    is a base. -/
  (isBase_exchange : Matroid.ExchangeProperty IsBase)
  /-- Every independent subset `I` of a set `X` for is contained in a maximal independent
    subset of `X`. -/
  (maximality : ∀ X, X ⊆ E → Matroid.ExistsMaximalSubsetProperty Indep X)
  /-- Every base is contained in the ground set. -/
  (subset_ground : ∀ B, IsBase B → B ⊆ E)
\end{lstlisting}
The use of both independence and base axioms introduces redundancy, but it is straightforward to derive either set of axioms from the other. This makes it easier to relate equivalent matroid definitions to each other, and also makes the more commonly used independence and base axioms more readily available for other definitions. To that note, this definition is not intended to be used directly. Instead, it is expected that a user takes advantage of appropriate constructors for the set of matroid axioms they are using.

\subsection{Graphs}

Graph theory is even thornier to work with in Lean. The graph theory library is smaller than libraries for other mathematical theories, and is less developed for a lot of subfields. This is mainly because the number of possible graph theoretic definitions leads to a combinatorial explosion. We want directed graphs, multigraphs, directed multigraphs, simple graphs. Do we want hypergraphs? Do we want graphs on distinct vertex types? How do we avoid code duplication, i.e. how do we avoid having to define and re-prove the same concepts separately for every possible definition?



We provide a brief history of different graph definitions \footnote{We omit the discussion on \lstinline{Quiver} as it behaves as a skeleton for \lstinline{Category}, but it can be thought of as a generalization of \lstinline{Digraph}.} in \texttt{Mathlib} and the reasons behind them, and examine a common weakness. Then we provide the most recent definition and explain how it avoids this weakness. 

We first consider the original definition of a simple graph in \texttt{Mathlib} from around 2020, translated into \lstinline{Lean 4} syntax and with some tactics and notation omitted for readability. We also include the definition of a directed graph, added to \texttt{Mathlib} in 2024. These definitions were inspired by the existing algebra hierarchy, as well as the graph theory library of Rocq\cite{Doczkal2020}. 

\begin{lstlisting}
structure SimpleGraph (V : Type u) where
  Adj : V → V → Prop
  symm : Symmetric Adj
  loopless : Irreflexive Adj

structure Digraph (V : Type*) where
  Adj : V → V → Prop    
\end{lstlisting}

One common graph theoretic definition that we want is subgraphs, whether they are substructures of \lstinline{SimpleGraph}, \lstinline{Digraph}, etc. If we define a subgraph as a \lstinline{SimpleGraph} on a subtype of the vertex type, this is a non-starter, as we will constantly have to coerce vertices that belong to both types. For this reason, we define \lstinline{Subgraph G} on \lstinline{G : SimpleGraph V} in a manner that imitates the design pattern of \lstinline{Group} and \lstinline{Subgroup}.

\begin{lstlisting}
structure Subgraph {V : Type u} 
    (G : SimpleGraph V) where
  /-- Vertices of the subgraph -/
  verts : Set V
  Adj : V → V → Prop
  /-- Adjacency inherits from parent graph -/
  adj_sub : ∀ {v w : V}, Adj v w → G.Adj v w
  /-- Adjacency implies carrier set membership -/
  edge_vert : ∀ {v w : V}, Adj v w → 
    v ∈ verts
  symm : Symmetric Adj := by aesop_graph
\end{lstlisting}

This is a correct definition for subgraph, but now suppose we have some \lstinline{H : Subgraph G} of \lstinline{G : SimpleGraph V} for which we want to use a \lstinline{SimpleGraph} lemma (without having to reprove it for \lstinline{Subgraph}). If we coerce \lstinline{H} to a \lstinline{SimpleGraph}, we have to plug the carrier set \lstinline{verts} into the type parameter. This will coerce \lstinline{verts} into a subtype of \lstinline{V}, and we run into the type coercion problem all over again. 

This sort of type coercion is less of a problem in the \lstinline{Group} and \lstinline{Subgroup} context, but modifying the vertices and edges of a graph via deletion, edge contraction, edge subdivision, etc is very common in graph theoretic proofs. The \lstinline{verts} coercion to a type means that every time we want to delete a vertex, introduce a new vertex by subdividing an edge, contract an edge, etc, we create graphs on new vertex type, with a mess of coercion maps between all of them. 





This leads us to the conclusion that, in order for us to reason about subgraphs and minors of graphs, we need to define all graphs with respect to sets of their vertex type instead of defining them on the vertex type itself. This conclusion is not dissimilar to the observation we made about matroids being defined as sets in an ambient type. In the case of \lstinline{SimpleGraph}, we would modify the definition to resemble something like the following: 

\begin{lstlisting}
structure SimpleGraph (V : Type u) where
  /-- Vertex carrier set -/
  vertexSet : Set V
  Adj : V → V → Prop
  /-- Set membership condition -/
  adjMem : 
    ∀ ⦃x y⦄, Adj x y → x ∈ vertexSet
  symm : Symmetric Adj
  loopless : Irreflexive Adj
\end{lstlisting}
One might wonder why we did not start with carrier vertex sets for our definitions in the first place. Consider the following: if we have a graph defined on a set of vertices belonging to an ambient type, we naturally want the adjacency relation to be restricted only to the vertex set. This requires us to include constraint \lstinline{adjMem} in our definition, so any time we interact with the adjacency relation of a \lstinline{SimpleGraph}, we are forced to carry proofs that \lstinline{adjMem} is satisfied. However, since we concluded that the carrier set interfaces cleanly with \lstinline{Subgraph}s, we reason that friction we may encounter when our automated \lstinline{adjMem} proofs fail is a worthwhile tradeoff.

We would also like our graph theory definitions to interface nicely with our matroid theory definitions, and graphic matroids are generally defined on multigraphs. As of May 2025, the first definition for this purpose in \texttt{Mathlib} applies the observation about carrier sets to both vertex type \lstinline{α} and edge type \lstinline{β}:


\begin{lstlisting}
structure Graph (α β : Type*) where
  /-- The vertex set. -/
  vertexSet : Set α
  /-- The binary incidence predicate, stating that `x` and `y` are the ends of an edge `e`. -/
  IsLink : β → α → α → Prop
  /-- The edge set. -/
  edgeSet : Set β := {e | ∃ x y, IsLink e x y}
  /-- If `e` goes from `x` to `y`, it goes from `y` to `x`. -/
  isLink_symm : ∀ ⦃e⦄, e ∈ edgeSet → (Symmetric <| IsLink e)
  /-- An edge is incident with at most one pair of vertices. -/
  eq_or_eq_of_isLink_of_isLink : 
    ∀ ⦃e x y v w⦄, IsLink e x y → 
    IsLink e v w → x = v ∨ x = w
  /-- An edge `e` is incident to something if and only if `e` is in the edge set. -/
  edge_mem_iff_exists_isLink : 
    ∀ e, e ∈ edgeSet ↔ ∃ x y, IsLink e x y
  /-- If some edge `e` is incident to `x`, then `x ∈ V`. -/
  left_mem_of_isLink : 
    ∀ ⦃e x y⦄, IsLink e x y → x ∈ vertexSet
\end{lstlisting}
Though it was arrived at independently via observations from III-C about matroids, this definition resembles the definition of undirected (labeled) graphs in HOL \cite{Chou1994}. However, notable differences include the fact that the edge set is an argument to the definition in HOL, whereas the edge set is derived from the vertex set and \lstinline{IsLink} relation in our definition. The well-defined condition for edge endpoints is also constructed differently. Future work may include inter-library comparisons, and how logic systems of different proof assistants lead to different definitions.

We have addressed many of the questions asked at the beginning of this section, but not all. For the question about graphs on distinct vertex types, the use case is fairly limited, and for now the solution is to define the vertex type for graphs on distinct types as being made up of the disjoint sum of the types. As for hypergraphs, we would likely have to generalize the definition of \lstinline{Graph} in some way, perhaps by having the \lstinline{IsLink} relation defined as \lstinline{β → Set α → Prop}. This is still an open question.

\section{Observations from Computer Algebra}
Computer algebra libraries based on algebraic types face most of the same issues we have articulated for mathematics formalization systems.
These arise to varying extents in systems such as Axiom, with the Aldor programming language, Magma, SageMath, MacCaulay 2, CoCoa and the Maple Domains package.   As these systems are primarily concerned with computing values, the emphasis is somewhat different than in mathematics formaliztion so additional issues arise.   We illustrate some of these using Aldor.

\subsection{Thumbnail Sketch of Aldor}
The Aldor~\cite{AldorCAH} programming language was first introduced as $A^\sharp$~\cite{AldorISSAC} as an extension language to the Axiom~\cite{Axiom} computer algebra system.  It generalized an earlier language~\cite{Symsac81} used while the Scratchpad II research system evolved to become Axiom.   

\begin{figure}[t]
\begin{footnotesize}
\begin{verbatim}
NNI ==> NonNegativeInteger;

define AbelianMonoid: Category ==
    Set with { +: (%, %) -> %;  0: % };
define AbelianGroup: Category ==
    AbelianMonoid with { -: % -> % };
define Ring: Category ==
    Join(AbelianGroup, Monoid) with
        Distributive(+, *);
define Module(R: Ring): Category ==
    Ring with *: (R, %) -> % };

UnivariatePolynomial(R: Ring): Module(R) with {
    monom:    (R, NNI) -> %;
    zero?:    % -> Boolean;
    degree:   % -> NNI;
    lc:       % -> R;  -- Leading coefficient
    reductum: % -> %;  -- Drop leading monomial
    if R has Commutative(*) then Commutative(*);
} == add {
    Rep == List Record(coeff: R, degree: NNI);
    0 == per [];
    1 == per [[1,0]];
    monom(coeff: R, degree: NNI) ==
        per [ [coeff, degree] ];
    zero?(p: %) ==
        null? rep p;
    degree(p: %) ==
        if zero? p then 0 else first(rep p).degree;
    lc(p: %) ==
        if zero? p then 0 else first(rep p).coeff;
    reductum(p: %) ==
        if zero? p then 0 else per cdr rep p;
    (r: R) * (p: T) ==
        if zero? r then 0
        else per map(t +-> r * t.coeff, rep p);
    (p: %) + (q: %) == {
        zero? p => q;
        zero? q => p;
        dp := degree p; dq := degree q;
        if dp > dq then
            per cons([lc p, dp], rep(reductum p + q))
        else if dp < dq then
            per cons([lc q, dq], rep(p + reductum p))
        else
            per cons([lc p + lc q, dp],
                     rep(reductum p + reductum q))
    }
}
\end{verbatim}
\end{footnotesize}
\vspace{-1.5\baselineskip}
\caption{A simplified Aldor example}
\label{fig:AldorEg}
\vspace{-\baselineskip}
\end{figure}
In Aldor, types, including dependent mapping types and dependent product types, are first class values.
The language provides mechanisms for creating abstract data types known as \textit{domains}, which provide the data representation for values.   
Every value belongs to a unique domain, but may belong to any number of subtypes of that domain.
The data representation of domains themselves (as values) is given by the domain \texttt{Domain}.
Subtypes of \texttt{Domain} are known as \textit{categories} and the language provides declarative features whereby categories may require domains to have specified operations or properties.
Typically, in library code, domains are produced by functions of dependent type and operations provided by the domain values are used to construct algebraic objects.

A simplified example is given in Figure~\ref{fig:AldorEg}.
To read this example we observe that 
names may be associated with values using either \verb+==+ or \verb+:=+ for constant or variable assignments respectively.
Constant assignments prefixed by the keyword \verb+define+  have their values visible in other compilation units, otherwise only the type of the constant is visible.
A function definition \verb+f(a: A): R == E+ gives \verb+f+ a value of type \verb+(a: A) -> R+, with \verb+R+ potentially depending on \verb+a+.
The basic operation to form categories is \verb+with+ and to form domains is \verb+add+.
Categories can specify subtypes of \texttt{Domain} as those domains exporting certain operations or belonging to specific other categories.
The name \verb+%+ denotes the type under construction, with an implicit fixed point operation applied.
Domains provide a representation type, given by \texttt{Rep}, and the operations \verb+rep+ and \verb+per+ perform coercions
$\text{\texttt{\%}} \rightarrow \text{\texttt{Rep}}$ and 
$\text{\texttt{Rep}} \rightarrow \text{\texttt{\%}}$ 
 respectively.
 In this example, 
the bracket operation \verb+[...]+ is used to construct \verb+Record+ and \verb+List+ values, and
both \verb+Module+ and \verb+UnivariatePolynomial+ are of dependent mapping types.

The example shows that within the code for the polynomial operations, it may be relied upon that the coefficients of polynomial values belong to the parameter ring \verb+R+ and therefore the arithmetic operations on \verb+R+ will be available.   This is known at compile time, even though the particular value of \verb+R+ is not known.  

There are some aspects to the Aldor type system that were pragmatic at the time of its development:
Type equivalence is determined by syntactic equality of type expressions.  That way, type expressions may involve user-defined functions without interpretation.
It is not checked that types are inhabited, nor that implementations satisfy declared algebraic identities (\textit{e.g.} that a supposedly commutative multiplication actually is).
Likewise it is not verified that programs terminate.   There is not a hierarchy of universes --- all types are at the same level and Russell's paradox is avoided by limiting the predicates available.

\subsection{Additional Issues Arising}

\paragraph{Language {\itshape vs} Library}  
It is desirable to separate a core programming language definition, which is fixed, from the definitions in an open-ended library.   
Consequently, in larger applications one often sees extensions of basic types (such as \verb+String+) to support the functions of various libraries and this then requires (implicit or explicit) coercions between the various derived types (\textit{e.g.} \verb+CorbaString+, \verb+SqlString+...).   This can arise because the derived objects require additional data fields. In the computer algebra setting, it is quite common to want to ascribe additional mathematical properties to values without requiring any modification of the representation.    For example, once the type \verb+Integer+ is defined, it may be desired to say that it satisfies the properties of the categories \verb+EuclideanDomain+ or \verb+PartialDifferentialRing+, defined later defined in libraries.   
This requirement becomes particularly acute if langauge-defined categories require mention of specific types, \textit{e.g.} \verb+Boolean+ values for signatures of relations.
To do this, Aldor provides the ability to declare domains to satisfy additional categories with so-called ``post facto'' extensions~\cite{AldorPostFacto}.  This can be simulated in programming languages such as \verb_C++_ or Rust with ``traits''.

\paragraph{Compilation}
In contrast to mathematical libraries of proof systems, where a proofs are often ``once and done'', computer algebra code is expected to be executed for multiple computations so efficiency of execution is of paramount importance.
When a compiled programming language is used, it is desirable that independently written libraries can be compiled separately, not requiring re-compilation of each use.  
This imposes constraints on the programming language, leading to such things as using syntactic equality of type expressions for type equivalence, the requirement for explicit marking of cross-unit value visibility (\verb+define+, described above), and limiting the predicates available on type expressions.

\paragraph{Subtyping and Mutability}
There continues to be an active question whether mutability of data is required to develop sufficiently efficient computer algebra code.  Most present systems do provide at least certain mutable data structures, and this has consequences on subtyping.   To see this, suppose there is a type of read-only vectors \verb+Vec T+.   Then if \verb+S+ is a subtype of $T$, with $e \in S \Rightarrow e \in T$, it can make sense to have covariant subtyping so \verb+Vec S+ be a subtype of \verb+Vec T+.   Every operation on a value of type \verb+Vec T+ would be valid on a value of type \verb+Vec S+, \textit{e.g.} element selection \verb+elt: (Vec T, NNI) -> T+ could be applied to a value of type \verb+Vec S+.     On the other hand, suppose the vectors were mutable.  In this case, the operation \verb+setelt!: (Vec T, NNI, T) -> T+ to set an element precludes having \verb+Vec S+ be a subtype of \verb+Vec T+.  To see this, suppose \verb+S = NNI+ is a subtype of \verb+T = Integer+ and \verb+vnni+ is of type \verb+Vector NNI+.  Then \verb+setelt!(vnni, 3, -1)+ would modify \verb+vnni+ to have a negative element.   For this reason, subtyping of parameterized types may need to be dealt with explicitly, such as in the Scala programming language.

\paragraph{Algebraic Structure of Subtypes}
In mathematics, it is often useful to derive certain types as generalizations of more basic types, \textit{e.g.} to think of the integers as the closure under subtraction of the natural numbers.  
When building libraries of executable code, however, it is often more practical to build a general type and then define subtypes based on a membership predicate, for example to have natural numbers to be the integers satisfying a test for non-negativity.    In this situation, the subtype may have more or less structure than the parent type.   For example, while the integers form a ring, the natural numbers as a subtype of the integers, do not.   As another example, restrictint the complex numbers to the unit circle adds the structure of a multiplicative group (with complex $\times$)  but loses the field structure (with complex $\times$ and $+$).

\paragraph{Binary Operations}
In object-oriented programming languages, it is usual to have an implicit first argument to operations and for derived types to subtype on this implicit argument.     For example, suppose the object class \verb+Polygon+ has \verb+perimeter+ operation the signature of which we denote as \verb+perimeter: /Polygon/ -> double+.  Here we write \verb+/Polygon/+ for the implicit first argument.   The implementation of the \verb+perimeter+ operation might add the lengths of all the sides of a data object.   
Now suppose we have a derived class, \verb+Square+.  In this case, there would be the inherited operation \verb+perimiter: /Square/ -> double+ and the implementation in \verb+Polygon+ would still be applicable, or a specialized implementation quadrupling the length of the first side could be provided.
Now consider the situation when the operation is binary, such as a law of composition (\textit{e.g.} $+$) or a relation (\textit{e.g.} $<$).   In this case, one wants both arguments to behave similarly.  This is one of the reasons that Aldor provides \verb+%+  as the fixed point of the type under construction.  A simple example of this occurs in providing an implementation of Gr\"obner bases.   Here, polynomials of $n$ variables may be viewed as univariate with exponent vectors of $n$ non-negative integers.  Then it is desired to imbue the exponent vectors with different orderings, such as lexicographic order or total degree order with reverse lexicographic tie-breaking.   In this situation, both arguments of the exponent order $<$ must be seen as extensions of the base type.
For this reason, some modern programming langauges have recently introduced the notion of \verb+thistype+ as similar to Aldor and Axiom's \verb+%+.

\spacetune{\balance}
\section{Advice}

It is difficult to create a rule set for formalizing mathematical definitions, but in this paper we hope to provide some general bits of advice\footnote{Supplementing and/or summarizing ideas from texts such as \cite{tpil4} or \cite{avigad2025mathematics}}. 

It is difficult to recognize the needed properties of a definition without experimenting, so it is better to pick a definition and "get the ball rolling," so to speak. In the case of the graph theory library described in III-D, this was a necessary step on the way to developing a more appropriate definition.

However, one must avoid prematurely committing to an approach, and should write with the intention of frequently refactoring or adjusting definitions. Developing a mathematical hierarchy requires a strong understanding of the math as well as an openness to creative reframings of the definitions. This is what allowed the authors of the continuous functional calculus library in III-A to repeatedly generalize their definitions and facilitate Lean's automated inference system.

When working with definitions intended for a mathematical hierarchy (as is typically the case for formalizations in \texttt{Mathlib}), a careful consideration of generalizations or future work can lessen the amount of refactoring we must perform later. This often involves careful weakening or division of conditions, as well as the addition of redundant conditions to ensure correctness and ease of use. Examples of this can be found throughout all of the cases we have examined, such as continuous functional calculus being extended to other domains in III-A, or allowing infinite matroids in III-C. 

Lastly, we often require layers of definitions that build on each other instead of just one definition. This can happen in the form of "junk value" functions as is the case in III-A, or constructors for matroids derived via different axioms in III-C.
In the computer algebra  context, having a  late-binding  fixed-point type concept (\textit{e.g.}  \texttt{\%} or \texttt{thistype} )  greatly assists elegant construction of algebraic hierarchies where internal laws of composition are important.

\section{Conclusions}
We have explored a range of issues that arise in defining non-trivial mathematical software libraries.  These have been examined both from the point of view of formal proof systems and computer algebra systems.  Related problems arise from the different uses of dependent types, although the focus and methods to handle them differ in priority.

\bibliographystyle{IEEEtran}
\IfFileExists{IfExistsUseBBL.tex}{%

}{%
\bibliography{citations.bib}
}
\end{document}